# Comparative Study on Reliability Estimation Using Monte Carlo Simulation with Application to Cylindrical Pressure Vessel

Abolfazl Zolfaghari[1,*], Maryam Khoshkhoo[2]

[1]*Department of Mechanical Engineering, Tennessee Tech University, TN, USA*
[2]*School of Biomedical Engineering, Babol Noshirvani university of technology, Babol, Iran;*
*\*ab.zolfaghari.abbasghaleh@ gmail.com*

One of the methods to design products is reliability-based design, in which failure probability is usually used instead of safety factors. In the technique, it should not be less than a predetermined value. Choosing the proper design criterion is a challenging problem for designers who are dealing with the technique, particularly, when there are various criteria. One of these kinds of products is cylindrical pressure vessel which diverse criteria proposed in the literature to calculate the burst pressure as a start point of the design. In this paper, we are going to evaluate and compare the performances of various burst pressure criteria in estimating failure probability which is used for a sample pressure vessel. For each criterion, Monte Carlo simulation has been employed to calculate the probability of failure due to variations related to major design variables. The design parameters include material properties and operating pressure. First, the effects of variations in standard deviations of the design variables on the calculated burst probabilities have been determined by standard deviation analysis. Then, sensitivity analyses were carried out to assess the sensitivity of each burst pressure criterion against changes in the magnitude of design variables.

## 1. Introduction

Pressure vessels is a vital component of industries dealing with pressurized gas or liquid when it is necessary to store it. They are usually spherical or cylindrical in shape, but the cylindrical type is normally chosen because of its ease for manufacturing. One of the main key factors in safe design, maintenance and operation of the vessel is its integrity, which it becomes even more crucial as working condition of the vessel is tough such as high temperature or pressure, or it contains hazardous materials. Traditional techniques of mechanical design of the vessels are based on safety factors, which are based on specifying safety factors to the component [1], [2]. Reliability based design is a modern method of design. In this approach, failure probability is used instead of the safety factors in the traditional technique. Reliability based design of products could be achieved by specifying the proper level of reliability in the system to account for all uncertainties and variabilities inherited by manufacturing process or working conditions. Reliability analysis can show the impact of each design element on the failure probability of the total system [3]-[7].

In reliability-based design, reliability estimation is the most important factor. A major challenge in reliability-based techniques is the need for large number of data to compute the failure probability of the system. Such data may be gathered experimentally which can lead to unacceptable costs. As an alternative, Monte Carlo Simulation (MCS) is a numerical iterative technique based on generating samples from a probability distribution. More recently, the authors of [8] proposed a scenario-generation method based on MCS which provide efficient sampling and is suitable for applications with computational challenges such as mixed-integer programming. In reliability prediction of system, fraction of generated samples which result to failure to total number of generated samples is the reliability value. It becomes more popular when dealing with problems containing non-linear equations and different types of random numbers distribution. The literature in applying MCS for reliability analysis of the systems is rather vast. From recent instances, based on Monte Carlo Simulation and nonlinear finite element analysis (FEA), a model has been proposed and utilized to analyze reliability of structures [9]. In another application, Cardoso et al [10], made effort to develop a method in structural reliability investigation using MCS and neural networks. Filababadi and Bagheri proposed a novel optimization-based reliability approach in [11] that takes into account structural features and topological components to respond to unexpected events. Kozak and Liel [12] evaluated the reliability of two kinds of roof structures carrying snow weight using MCS. Lotfollahi et al. proposed a system based on reliability using MSC connected to the nonlinear finite element program to assess the fragility of a special structure [13]. Mahmoud and Riveros [14] used a hybrid of MCS and FEM to assess fatigue reliability of a specific type of stiffened panel applying in ship building. Daryani and Mohamad [15] applied the simulation to assess the reliability of some kinds of retaining walls. In another work, the method was applied to evaluate impacts of structural and aerodynamic uncertainties on the reliability of bridge in [16]. In a similar way, the reliability of a hull girder harnessing Monte Carlo simulation technique analyzed in [17] and a methodology founded on reliability for drilling shafts subjected to axial loads recommended in [18].

In pressure vessels and pipelines engineering, some recent researches have been listed in the following. Al-Amin and Zhou evaluated the reliability of pipelines under the corrosion hiring combination of simple Monte Carlo and Markov Chain Monte Carlo [19]. Monte Carlo simulation was employed as a great tool to analyze the reliability of a pipeline in China introduced in [20]. Utilizing MCS related to an analytical model, Bouhafs et al. [21] have studied the probability of failing in the thick composite pipes under internal pressure. The results revealed that the pipe thickness and changing in the internal pressure are the principal factors affecting the scatter of stress distribution. Zhang et al. [22] proposed a probabilistic approach to evaluate the integrity of pipelines carrying corrosion defect. Then, the failure probability of the pipelines was predicted by coupling the mentioned method and MCS. According to the findings, the most effective parameters in larger effect on the obtained probability included operating pressure and defect length. Altamura and Beretta presented the probabilistic model and the reliability evaluation of tubular mechanical components carrying some flaws subjected to fatigue loading with variation in internal pressure using MCS [23]. Zhou et al. [24] have proposed a technique that applied MCS to assess time-dependent system reliability of the corroding pipeline part holding multiple spatially correlated active corrosion defects.

A major challenge in reliability-based design of systems is selection a technique to determine level of reliability of the system, which it might be confusing for designers. The pressure vessel is a good example of these kinds of systems due to number of formulas have been proposed and utilized to determine burst pressure [25], which is uncertain and mainly depends on the design of the vessels. Such uncertainties in burst pressure can lead to serious reliability and economic issues [24], [25]. Recent predictive methods such as particle swarm optimization [26], fuzzy-based models [27], feed neural networks [28], and artificial neural networks [29] have shown to be efficient in various applications.

Alternatively, uncertainty management was proposed in the literature to address uncertainty in the models. Among various uncertainty tackling models proposed in the literature, we have found the most recent proposed approaches, i.e., [30] and [31], more suitable to this application due to the flexibility to adjust the uncertainty of burst pressure based on historical data and system budgets. The merits of such qualitative approaches have been theoretically and numerically demonstrated in various applications such as renewable energy management and power systems in [30], [32], and [33]. However, these approaches require an optimizer in the system to collaborate with the designer which requires budgets for protecting the system under severe cases. This can be of interest in the future research to utilize the use of such optimization techniques in experimental reliability studies for parameter tuning.

Contradictory to uncertainty modeling approaches, we can use efficient predictive methods for some applications such as burst pressure. In this paper, we investigate the selection of the criterion that impacts the reliability of pressure vessels and implement five predictive methods to predict burst pressure of unflawed thin walled pressure vessel. Such predictive methods are studied through one illustrative sample of a pressure vessel with known statistical parameters regarding design variables. The design variables operate pressure and material properties including yield and ultimate strength. To achieve the aim, we develop a procedure that is implemented based on Monte Carlo simulation to predict the failure of probabilities corresponding to each criterion. Then, for each criterion, the changing effects in the mean and standard deviation of the design variables on the forecasted failure probability are studied. At the end, sensitivity analysis is carried out to analyze the selection criteria as well.

## 2. Burst in Pressure Vessels

Depending on working condition and level of required safety, several types of failure such as small cracks, excessive leakage, and fracture/catastrophic rupture are defined [34]. However, catastrophic rupture is the most serious failure mode of pressure vessels because it could occur suddenly in the vessels without any signs like leakage, which may result in damaging properties or people lives. When internal pressure of the vessel increases, at specific value, it abruptly ruptures since it is beyond load capacity of the vessel. As a result, prediction of the pressure is crucial not only in design based on safety factors but also reliability-based design, which is probability of failure basis. Accurate estimate of the pressure leads to lower cost of material without losing the safety. To have more accurate the pressure predictor, a numerous formulas and models have been introduced in the literature [35]-[44]. As mentioned above, the criteria used in this paper are those mostly developed for predicting the burst pressure of unflawed thin walled cylindrical pressure vessels manufactured by mild steel. Descriptions are presented in Table 1.

## 3. Reliability Estimation and Monte Carlo Simulation

Probability of failure and reliability are generally calculated by establishing a limit-state function, $g(X_1, X_2, \ldots, X_n)$. The instance where $g(X_1, X_2, \ldots, X_n) \leq 0$ shows failure state of the system. Hence, the probability of failure is calculated as bellow. In what follows, we first present the nomenclature and then the related discussion and equations.

---

Nomenclature

| | |
|---|---|
| $D_i$ | Inner diameter (m) |
| $D_o$ | Outer diameter (m) |
| $C_v$ | Coefficient of Variation |
| e | Base of natural logarithm |
| $F_X(x)$ | Probability density function (pdf) of X |
| $g(X_1,\ldots,X_n)$ | Limit state (or failure) function |
| i | The number of iteration |
| j | Number of trial result in failure |
| k | Algorithm counter |
| m | Total number of trials |
| n | The number of variables |
| P | Probability function |
| $P_b$ | Burst pressure (Pa) |
| $P_f$ | Probability of the failure |
| $P_o$ | Operating internal pressure (Pa) |
| r | Strain-hardening exponent |
| $R_f$ | Reliability of the system |
| X | Random variable (basic variable) |
| $(x_1, x_2, \ldots, x_n)$ | Randomly generated values for $X_1, X_2, \ldots, X_n$ |
| α | Mean value |
| μ | Standard deviation |
| $α_i$ | Sensitivity coefficient |
| $σ_u$ | Ultimate strength (Pa) |
| $σ_y$ | Yield strength (Pa) |

Table 1. Burst pressure calculation criteria

| Criterion | Features | Formula |
|---|---|---|
| Faupel [37] | Empirical, general | $P_b = \dfrac{2}{\sqrt{3}}\sigma_y(2-\dfrac{\sigma_y}{\sigma_u})\ln(\dfrac{D_o}{D_i})$ |
| Svensson [38] | Theoretical, thin-walled | $P_b = \sigma_u(\dfrac{0.25}{n+0.227})(\dfrac{e}{n})^r(\ln(\dfrac{D_o}{D_i}))$ |
|  |  | where $r = 0.224(\dfrac{\sigma_u}{\sigma_y}-1)^{0.604}$ |
| Christopher et al [39] | Theoretical, thin-walled | $P_b = \dfrac{2}{(\sqrt{3})^{r+1}}\sigma_u\dfrac{(D_o-D_i)}{D_i}$ |
| Modified Faupel (Zheng et al. [40]) | Experimental, mild steel | $P_b = 13.21\sigma_y(\sigma_y/\sigma_u)^4\ln(\dfrac{D_o}{D_i})$ |
| Modified Faupel (Barbin et al. [41]) | Numerical, ductile steel | $P_b = \dfrac{2}{\sqrt{3}}\sigma_y\{1+0.65(1-\dfrac{\sigma_y}{\sigma_u})\}\ln(\dfrac{D_o}{D_i})$ |

$$P_f = P[g(X_1, X_2, ..., X_n) \leq 0] \tag{1}$$

$$P_f = \int\int...\int_{g(X_1,X_2,...,X_n)\leq 0} f_{X_1,X_2,...,X_n}(x_1, x_2, ..., x_n)dx_1 dx_2...dx_n \tag{2}$$

$$R_f = 1 - P_f \tag{3}$$

For multi-random variables problem or complex limit-state, determining the value of failure probability using the above relationship may seem complicated or a costly task. In such situations, numerical or analytical methods may be employed. In this regard, simulation-based methods, such as Monte Carlo simulation, are widely used to match the problem of estimating the failure probability. It becomes more popular when dealing with problems containing non-linear equations and different types of random number distribution. As an iterative procedure, it is based on random data generating within system's feasible ranges and assessing the failure state of the system. The ratio of the number of failed instances to the total number of trials converges to the probability of the failure when the number of iterations is increased and becomes large enough. Then, increasing in the number will not affect the probability of the failure. Thus, the probability of the failure would be defined by means of this method as follows.

$$P_f = \dfrac{1}{m}\sum_{i=1}^{m} I(X_1, X_2, ..., X_n) \tag{4}$$

$$I(X_1, X_2, ..., X_n) = \begin{cases} 1 & \text{if} \quad g(X_1, X_2,...X_n) \leq 0 \\ 0 & \text{if} \quad g(X_1, X_2,...X_n) > 0 \end{cases} \tag{5}$$

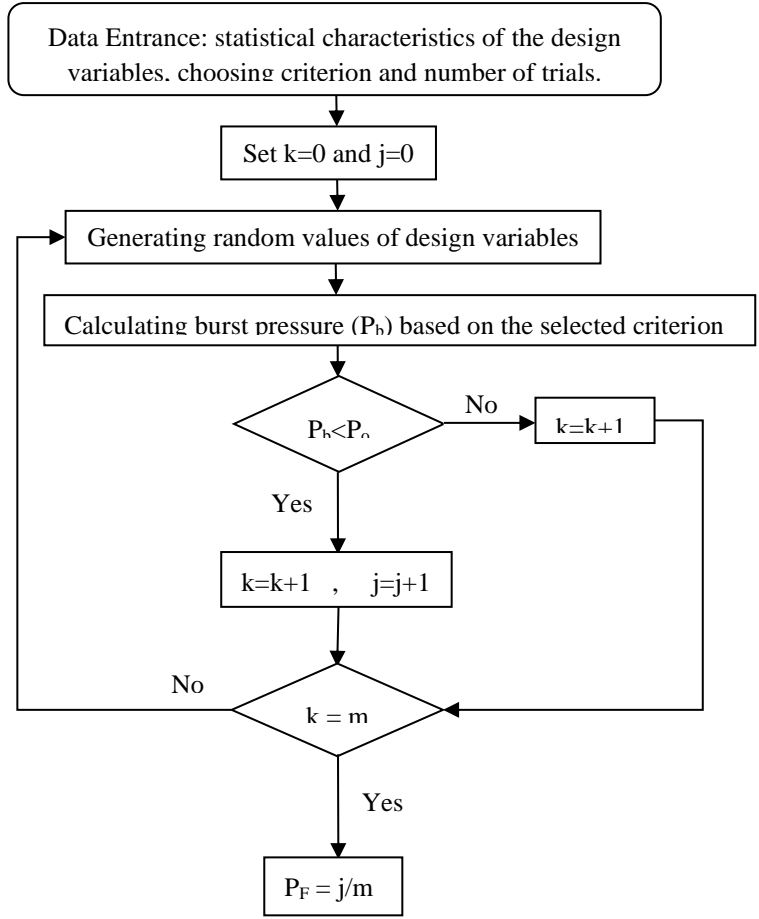

Fig. 1. Flowchart of the procedure in estimating the probability of failure based on MCS

For a pressure vessel that has uncertainties in material and geometric properties as well as variations in internal pressures, the corresponding limit-state function is defined as bellow:

$$g(X_1, X_2, ..., X_n) = P_b - P_o \tag{6}$$

As illustrated in Fig. 1 based on MCS, the failure probability of a pressure vessel is determined by using an iterative process developed in the research. At each iteration, values of design variables are statistically generated. Then, for the under-consideration criterion, the burst pressure is calculated and compared with the operating pressure. If the operating pressure is larger than the calculated burst pressure, the number of failed instances would be increased by one degree. The procedure is repeated until we reach the termination condition, the number of simulation trials.

## 4. Case Study

According to above explanation, in the reliability-based design of the pressure vessels, choosing the proper burst pressure criterion, as a design criterion, is also can affect the design safety. In the research, one sample of a pressure vessel is shown with normally distributed design variables. Hence, the proposed procedure was coded and executed in MATLAB software. The statistical description of the design variables and the geometrical characteristics of the vessel are listed in Table 2. According to the exercises, for all of criteria, the number of simulation runs should be around $10^5$ or larger, as illustrated in Fig. 2. Therefore, for each criterion, the Monte Carlo simulation was run $10^6$ times. The predicted failure probabilities are illustrated in Table 3.

Later, the effects of modification in the standard deviations and the mean values are evaluated and compared. It is due to possible alterations in working or manufacturing conditions, and the normally distributed design variables on the estimated probability of failure for each criterion. This would reveal how failure probability varies in modifications of the standard deviation values in design variables. Moreover, for all given criteria, sensitivity analysis, which shows the effects of changes in the mean values of the design variables on the probability of failure, could determine which one has greater results on the probability of failure.

Table 2. Statistical description of design variables and geometrical characteristics of the vessel

| Design variables | Mean value | Standard deviation |
|---|---|---|
| $P_o$ | 13 Mpa | 1 MPa |
| $\sigma_y$ | 235 MPa | 10 MPa |
| $\sigma_u$ | 375 MPa | 12 MPa |
| $D_o$ | 1000 mm | 0.5 mm |
| $D_i$ | 960 mm | 0.5 mm |

Table 3. Predicted failure probabilities

| Criterion | Predicted failure probabilities | Estimated burst pressure (MPa) | Average computational time (s) |
|---|---|---|---|
| Faupel [42] | 0.024 | 15.21 | 45.3 |
| Svensson [43] | 0.012 | 15.51 | 45.4 |
| Christopher et al. [44] | 0.001 | 16.49 | 45.9 |
| Modified Faupel (Zheng et al.) [45] | 0.06 | 19.54 | 45.2 |
| Modified Faupel (Barbin et al.) [46] | 0.248 | 13.77 | 45.4 |

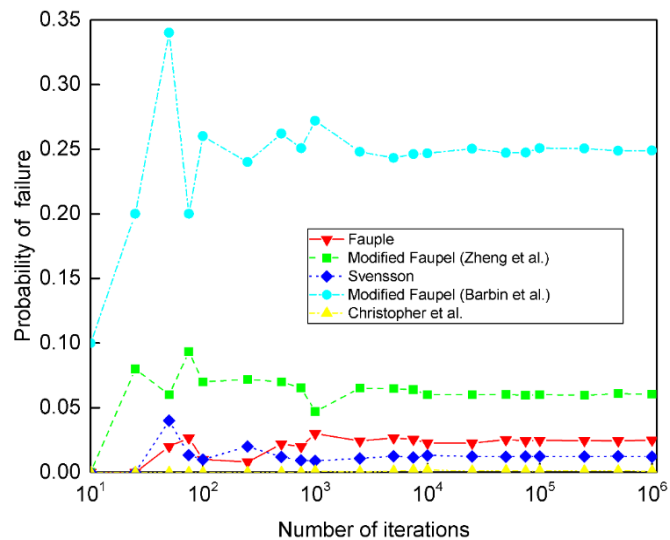

Fig. 2. The Convergence curves of Monte Carlo simulation

The results indicate the predicted failure probabilities largely fluctuate from 0.001 in Christopher et el. criterion to 0.248 in Modified Faupel (Barbin et al.). Hence, the same value specified to the system reliability for two different criteria, may not be resulted for sufficient, safe vessel design, while it is a conservative design according to another one.

The estimated burst pressures listed in Table 3 show that there is only one meaningful relationship, a lower burst pressure results to a higher failure probability between estimated burst pressure and corresponding failure probabilities in Modified Faupel (Barbin et al.) criterion. The predicted burst pressure obtained by applying the mean values of design variables and geometrical characteristics of the vessel. In the terms of computational time, all computational times are almost equal and there are no relationships between these times and iteration numbers of requirement which seems needed in finding failure probabilities based on Fig. 2, for speed convergence.

**4.1.** *Standard deviation analysis*

For specific ranges of design variables, changing effects in the standard deviation on the probability of failure are studied using graphs exhibiting the predicted failure probabilities against the standard deviation of the following variables.

For the range of 0.25 to 3 in the standard deviation, as illustrated in Fig. 3, the minimum alterations in the failure probability belongs to Modified Faupel (Zneng et al.). The corresponding failure probability increased from 0.054 to 0.101 due to increasing in the operating pressure in the range. On the other hand, the maximum affective criterion is Modified Faupel (Barbin et al.) caused by growing the probability from 0.083 to 0.403 regarding the pressure enhancement in the range.

4.1.1. *Operating pressure*

For the range of 0.25 to 3 MPa in the standard deviation, as illustrated in Fig. 3, the minimum alterations in the failure probability belongs to Modified Faupel (Zneng et al.). The corresponding failure probability increased from 0.054 to 0.101 due to increasing in the operating pressure in the range. On the other hand, the maximum affective criterion is Modified Faupel (Barbin et al.) caused by growing the probability from 0.083 to 0.403 regarding the pressure enhancement in the range.

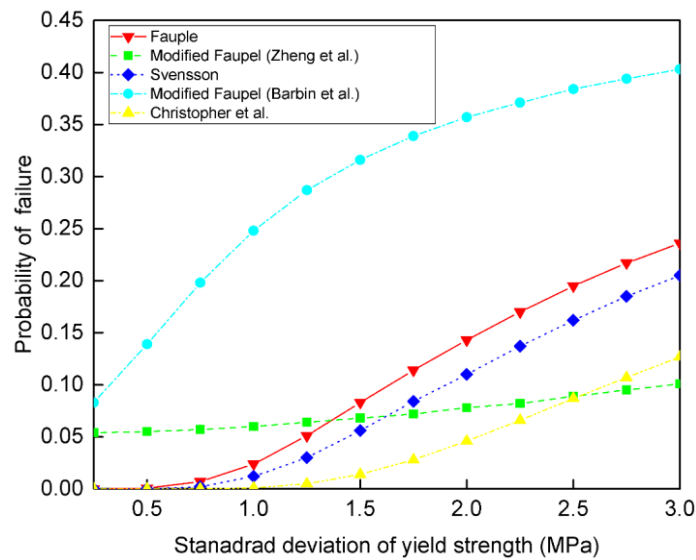

Fig. 3. Standard deviation analysis of operating pressure

### 4.1.2. *Yield and ultimate strength*

According to Fig. 4 and Fig. 5, for all criteria, trends of both describing the predicted probability of failure versus standard deviation of yield and ultimate strength are approximately the same. The obtained probability employing Modified Faupel (Zneng et al.) is the most sensible one to the standard deviation of yield and ultimate strength. For this criterion, the probability is growing from 0.003 to 0.227 in yield strength, as well as between 0.04 and 0.112 in ultimate strength when the standard deviation is increasing from 2 to 24 MPa. The changes in the estimated probability are not considerable for others as the mentioned criterion.

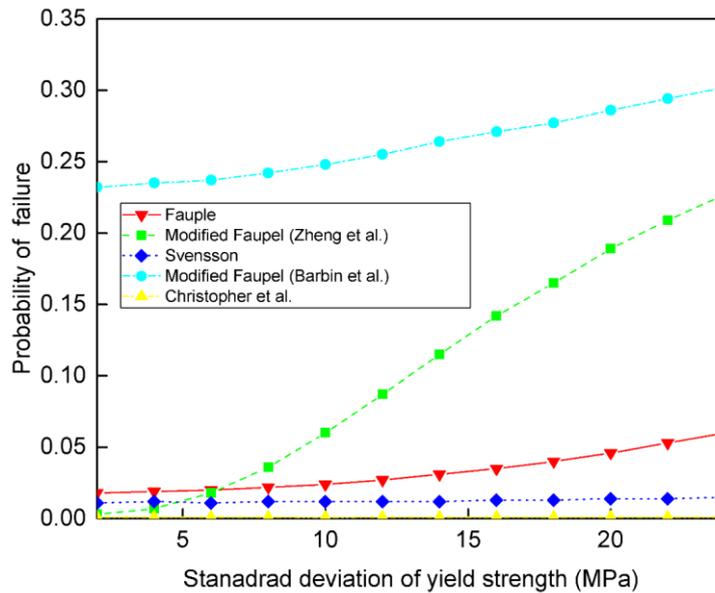

Fig. 4. Standard deviation analysis of yield strength

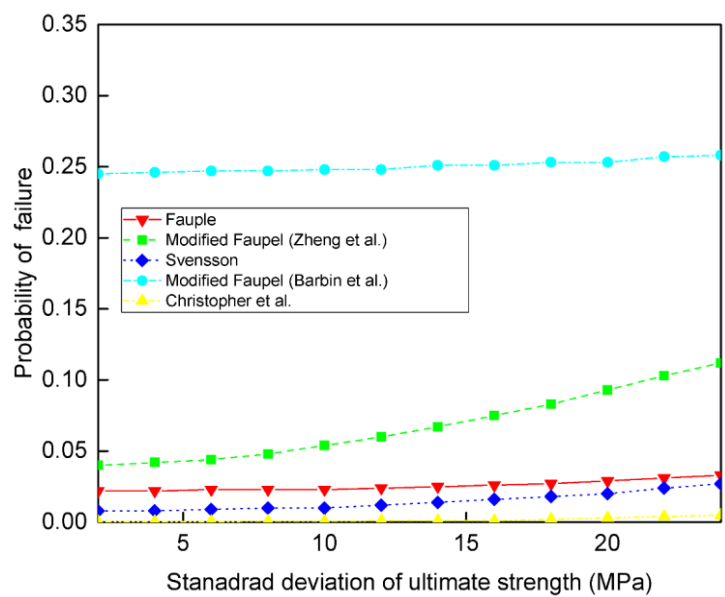

Fig. 5. Standard deviation analysis of ultimate strength

**4.2. *Sensitivity analysis***

Sensitivity analysis is a method to study the influence of modification in one or more input variables on the system outputs. One of the simple techniques in the method is, first applying derivative, a mathematical equation, and then finding sensitivity coefficient (αi) relating to each input factor. Factors with higher co-efficiency can affect the output more than the lower ones. The explained coefficient is determined by using Equation (7) or simplified form of Equation (8). It is obvious, in Equation (8), the coefficient is calculated, for each input parameter, by making a slight change in the parameter and observing the system effects. In the research, to assess how the design variables could affect the predicted probability of failure (POF), increments in coefficient of variation (COV), Cv, shown as Equation (9), and calculating the related sensitivity coefficient is used for each design variable. Based on the trials done before, 0.001 is a good option for the increment. The results of the analysis are shown in Table 4 and Table 5 as well as bar diagrams in Fig. 6 to Fig. 8.

$$\alpha_i = \frac{\partial P_f(X_1, X_2, ..., X_n)}{\partial X_i} \tag{7}$$

$$\alpha_i \approx \frac{P_f(X_1, X_2, ..., X_i + \Delta X_i, ..., X_n)}{\Delta X_i} - \frac{P_f(X_1, X_2, ..., X_n)}{\Delta X_i} \tag{8}$$

$$C_v = \frac{\sigma}{\mu} \tag{9}$$

By the data presented in Tables 4 and 5, sensitivity coefficient in operating pressure, is significantly larger than other coefficients relating to yield and ultimate strength for all the criteria. Thus, dominant factor in failing the vessel is operating pressure and must be controlled with more restricts. However, there is no exact rule demonstrating the sensitive system to yield strength or ultimate. For some certain criteria, such as Faupel and Modified Faupel (Barbin et al.), the system is more sensitive for ultimate strength than yield. Nevertheless, in case of the remaining criteria, the yield strength is more effective in the system.

Table 4. Sensitivity analysis (Faupel, Svensson, Christopher et al.)

| Design variables | POF | COV increment | POF increment | Sensitivity coefficient |
|---|---|---|---|---|
| Operating pressure | (0.024, 0.012, 0.001) | 0.001 | (0.01, 0.158, 0.054) | ($5.9 \times 10^{-2}$, $9.3 \times 10^{-1}$, $3.2 \times 10^{-1}$) |
| Yield strength | | 0.001 | (0.001, 0.003, 0.035) | ($1.8 \times 10^{-4}$, $5.4 \times 10^{-4}$, $6.3 \times 10^{-3}$) |
| Ultimate strength | | 0.001 | (0.01, 0.006, 0.068) | ($8.5 \times 10^{-4}$, $5.1 \times 10^{-4}$, $5.8 \times 10^{-3}$) |

Table 5. Sensitivity analysis (Modified Faupel (Zheng et al.), Modified Faupel (Barbin et al.))

| Design variables | POF | COV increment | POF increment | Sensitivity coefficient |
|---|---|---|---|---|
| Operating pressure | (0.06, 0.248) | 0.001 | (0.006, 0.052) | ($3.6 \times 10^{-2}$, $3.1 \times 10^{-1}$) |
| Yield strength | | 0.001 | (0.036, 0.013) | ($6.5 \times 10^{-3}$, $2.4 \times 10^{-3}$) |
| Ultimate strength | | 0.001 | (0.005, 0.05) | ($4.3 \times 10^{-4}$, $4.3 \times 10^{-3}$) |

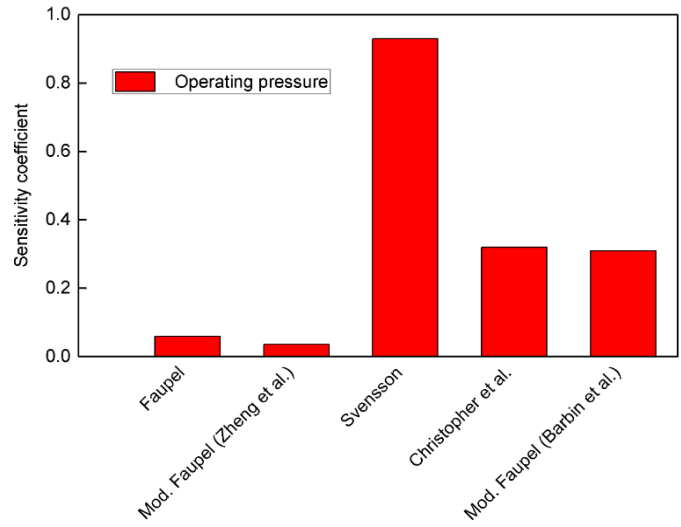

Fig. 6. Sensitivity analysis of operating pressure

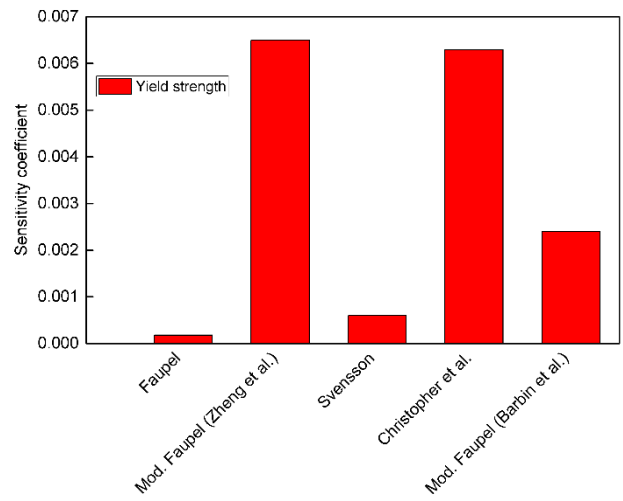

Fig. 7. Sensitivity analysis of yield strength

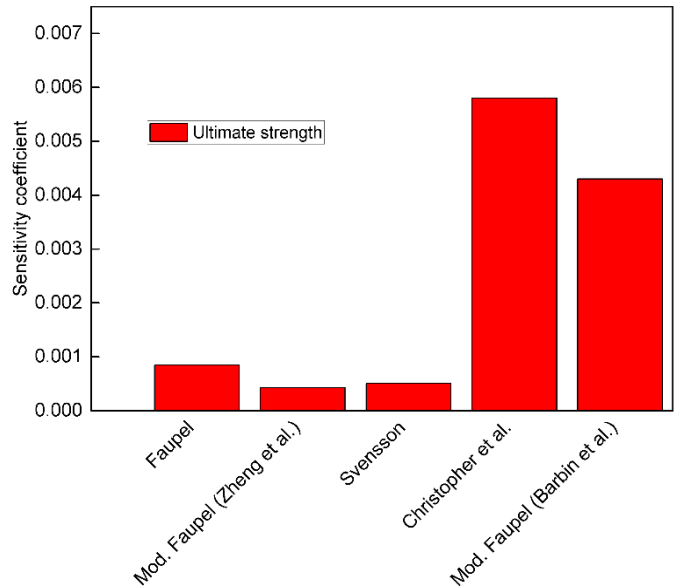

Fig. 8. Sensitivity analysis of ultimate strength

According to Fig. 6, Table 4, and Table 5, Modified Faupel (Zneng et al.) having sensitivity coefficient equals to 0.036 that is the least sensitive to variation in the mean value of operating pressure. However, Svensson is the most sensitive criteria because the calculated coefficient, by 0.93, is the largest.

The results of the sensitivity analysis of yield and ultimate strength have demonstrated as shown in Fig. 7 and Fig. 8 that Modified Faupel (Zneng et al.) with sensitivity coefficient equals to 0.0065 and Christopher et al. by 0.0058 are the most sensitive ones to mean value of yield and ultimate strength, respectively. Furthermore, Faupel is the least sensitive to yield strength; Svensson is the minimum sensitive to ultimate strength. Therefore, based on both the analysis, in choosing the minimum sensitive criterion to yield and ultimate strength, Modified Faupel (Zneng et al.) and Christopher et al. would not be the suitable criterion.

## 5. Conclusion

In this work, a probabilistic model has been introduced based on Monte Carlo simulation. Using the model, reliability of a sample of unflawed thin walled cylindrical pressure vessel was predicted and compared under diverse criteria. In the evaluation, design variables include operating pressure, yield and ultimate strength. The criteria include as Faupel, Svensson, Christopher et al., Modified Faupel (Zheng et al.), Modified Faupel (Barbin et al.). The results have shown that there are meaningful differences between the failure probabilities that are employing diverse criterion. It was varied between 0.001 and 0.248. Thus, choosing criterion has a big impact on the predicted probability of failure; the designer must be aware of that.

Then, two statistical tools, standard deviation and sensitivity analysis were employed to study the criteria deeply. The main findings using both tools are: (i) The operating pressure is the most effective factor in the failure comparing to the yield and ultimate strength. Thus, it should be controlled by paying more attention, (2) the failure probability has been obtained using Modified Faupel (Zneng et al.) is the least sensible criterion to standard deviation and the mean value of the operating pressure, (3) modified Faupel (Barbin et al.) and Svensson are the most effective to standard deviation and the mean value of the operating pressure, respectively, (4) the changes in standard deviation of yield and ultimate strength have the biggest impact on the probability of the failure estimated using Modified Faupel (Zneng et al.), and (5) modified Faupel (Zneng et al.) criterion is the main sensible criterion of the mean value of ultimate strength; the largest one to yield strength is Svensson.

In summary, there is no determined rule to choose criterion which is the least or more sensitized to the design variables. But generally, Faupel is not the least sensitive, it would be the least sensitive to standard deviation and the mean value in the all design variables. As a future research direction, one may need to investigate other suitable approaches for uncertainty, i.e., [30], [31], and implement them for dealing with the uncertainty of burst pressure. Such analysis can help users select the proper tools based on their aims and objectives and see the trade-off between uncertainty handling models, predictive models, and cost factors.

The future direction of this research is to generalize the proposed method to be used in various important applications such as microstereolithography [47, 48], printing [49], spatial wave-function switches [50-53]. Further, we would like to explore the possibility of building a new model upon the ground theories of [54-56] for voltage stability to integrate the simulation-related concepts in voltage control centers.

**References**

[1]. D. Annaratone, *Pressure Vessel Design*. Berlin Heidelberg: Springer-Verlag, 2007.

[2]. A. Zolfaghari, H. Razavi, and M. Izadi, "Optimum design of natural gas trunk line using simulated annealing algorithm," *IJOGCT*, vol. 1, no. 1, p. 1, 2020, doi: 10.1504/IJOGCT.2020.10029041.

[3]. "Probability Concepts in Engineering: Emphasis on Applications to Civil and Environmental Engineering, 2nd Edition | Wiley," *Wiley.com*. https://www.wiley.com/en-us/Probability+Concepts+in+Engineering%3A+Emphasis+on+Applications+to+Civil+and+Environmental+Engineering%2C+2nd+Edition-p-9780471720645 (accessed Apr. 28, 2020).

[4]. Y. Liu, H. Zhang, Y. Liu, Y. Deng, N. Jiang, and N. Lu, "Fatigue reliability assessment for orthotropic steel deck details under traffic flow and temperature loading," *Engineering Failure Analysis*, vol. 71, pp. 179–194, Jan. 2017, doi: 10.1016/j.engfailanal.2016.11.007.

[5]. A. Harding, "Reliability, Maintenance and Logistic Support: A Life Cycle Approach," *Journal of the Operational Research Society*, vol. 53, no. 12, pp. 1399–1400, Dec. 2002, doi: 10.1057/palgrave.jors.2601459.

[6]. M. Leimeister and A. Kolios, "A review of reliability-based methods for risk analysis and their application in the offshore wind industry," *Renewable and Sustainable Energy Reviews*, vol. 91, pp. 1065–1076, Aug. 2018, doi: 10.1016/j.rser.2018.04.004.

[7]. Charandabi, Sina E., Farnaz Ghashami, and Kamyar Kamyar. "US-China Tariff War: A Gravity Approach." *Business and Economic Research* 11, no. 3 (2021): 69-77.

[8]. E. Zio, "System Reliability and Risk Analysis by Monte Carlo Simulation," in *The Monte Carlo Simulation Method for System Reliability and Risk Analysis*, E. Zio, Ed. London: Springer, 2013, pp. 59–81.

[9]. Filabadi, Milad Dehghani, Afshin Asadi, Ramin Giahi, Ali Tahanpour Ardakani, and Ali Azadeh. "A New Stochastic Model for Bus Rapid Transit Scheduling with Uncertainty." *Future Transportation* 2, no. 1 (2022): 165-184.

[10]. J. B. Cardoso, J. R. de Almeida, J. M. Dias, and P. G. Coelho, "Structural reliability analysis using Monte Carlo simulation and neural networks," *Advances in Engineering Software*, vol. 39, no. 6, pp. 505–513, Jun. 2008, doi: 10.1016/j.advengsoft.2007.03.015.

[11]. Filabadi, Milad Dehghani, and Pegah Bagheri. "Robust-and-Cheap Framework for Network Resilience: A Novel Mixed-Integer Formulation and Solution Method." *arXiv preprint arXiv:2110.09694*, 2021. Available at: https://arxiv.org/pdf/2110.09694.pdf.

[12]. D. L. Kozak and A. B. Liel, "Reliability of steel roof structures under snow loads," *Structural Safety*, vol. 54, pp. 46–56, May 2015, doi: 10.1016/j.strusafe.2015.02.004.

[13]. M. Lotfollahi, M. Banazadeh, and M. M. Alinia, "Application of system reliability-based assessment for collapse fragility of braced moment-resisting frames," *The Structural Design of Tall and Special Buildings*, vol. 24, no. 11, pp. 757–778, 2015, doi: 10.1002/tal.1210.

[14]. H. Mahmoud and G. Riveros, "Fatigue reliability of a single stiffened ship hull panel," *Engineering Structures*, vol. 66, pp. 89–99, May 2014, doi: 10.1016/j.engstruct.2014.02.007.

[15]. K. E. Daryani and H. Mohamad, "System reliability-based analysis of cantilever retaining walls embedded in granular soils," *Georisk: Assessment and Management of Risk for Engineered Systems and Geohazards*, vol. 8, no. 3, pp. 192–201, Jul. 2014, doi: 10.1080/17499518.2014.937583.

[16]. T. Argentini, A. Pagani, D. Rocchi, and A. Zasso, "Monte Carlo analysis of total damping and flutter speed of a long span bridge: Effects of structural and aerodynamic uncertainties," *Journal of Wind Engineering and Industrial Aerodynamics*, vol. 128, pp. 90–104, May 2014, doi: 10.1016/j.jweia.2014.02.010.

[17]. B. Gaspar and C. Guedes Soares, "Hull girder reliability using a Monte Carlo based simulation method," *Probabilistic Engineering Mechanics*, vol. 31, pp. 65–75, Jan. 2013, doi: 10.1016/j.probengmech.2012.10.002.

[18]. H. Fan and R. Liang, "Reliability-based design of axially loaded drilled shafts using Monte Carlo method," *International Journal for Numerical and Analytical Methods in Geomechanics*, vol. 37, no. 14, pp. 2223–2238, 2013, doi: 10.1002/nag.2131.

[19]. M. Al-Amin and W. Zhou, "Evaluating the system reliability of corroding pipelines based on inspection data," *Structure and Infrastructure Engineering*, vol. 10, no. 9, pp. 1161–1175, Sep. 2014, doi: 10.1080/15732479.2013.793725.

[20]. K. Wen, J. Gong, B. Zhao, W. Zhang, and Z. Zhang, "The Reliability-Based Assessment of an In-Service X80 Natural Gas Pipeline in China," presented at the 2014 10th International Pipeline Conference, Dec. 2014, doi: 10.1115/IPC2014-33275.

[21]. M. Bouhafs, Z. Sereir, and A. Chateauneuf, "Probabilistic analysis of the mechanical response of thick composite pipes under internal pressure," *International Journal of Pressure Vessels and Piping*, vol. 95, pp. 7–15, Jul. 2012, doi: 10.1016/j.ijpvp.2012.05.001.

[22]. G. Zhang, J. Luo, X. Zhao, H. Zhang, L. Zhang, and Y. Zhang, "Research on probabilistic assessment method based on the corroded pipeline assessment criteria," *International Journal of Pressure Vessels and Piping*, vol. 95, pp. 1–6, Jul. 2012, doi: 10.1016/j.ijpvp.2011.12.001.

[23]. A. Altamura and S. Beretta, "Reliability assessment of hydraulic cylinders considering service loads and flaw distribution," *International Journal of Pressure Vessels and Piping*, vol. 98, pp. 76–88, Oct. 2012, doi: 10.1016/j.ijpvp.2012.07.006.

[24]. W. Zhou, H. P. Hong, and S. Zhang, "Impact of dependent stochastic defect growth on system reliability of corroding pipelines," *International Journal of Pressure Vessels and Piping*, vol. 96–97, pp. 68–77, Aug. 2012, doi: 10.1016/j.ijpvp.2012.06.005.

[25]. A. Zolfaghari and M. Izadi, "Burst Pressure Prediction of Cylindrical Vessels Using Artificial Neural Network," *J. Pressure Vessel Technol*, vol. 142, no. 3, Jun. 2020, doi: 10.1115/1.4045729.

[26]. Ghashami, Farnaz, Kamyar Kamyar, and S. Ali Riazi. "Prediction of Stock Market Index Using a Hybrid Technique of Artificial Neural Networks and Particle Swarm Optimization." *Applied Economics and Finance* 8, no. 3 (2021): 1-8.

[27]. Ghashami, Farnaz, and Kamyar Kamyar. "Performance Evaluation of ANFIS and GA-ANFIS for Predicting Stock Market Indices." *International Journal of Economics and Finance* 13, no. 7 (2021): 1-1.

[28]. Charandabi, Sina E., and Kamyar Kamyar. "Using A Feed Forward Neural Network Algorithm to Predict Prices of Multiple Cryptocurrencies." *European Journal of Business and Management Research* 6, no. 5 (2021): 15-19.

[29]. Charandabi, Sina E., and Kamyar Kamyar. "Prediction of Cryptocurrency Price Index Using Artificial Neural Networks: A Survey of the Literature." *European Journal of Business and Management Research* 6, no. 6 (2021): 17-20.

[30].   Filabadi, Milad Dehghani, and Houra Mahmoudzadeh. "Effective Budget of Uncertainty for Classes of Robust Optimization." *INFORMS Journal on Optimization, 2022, doi.org/10.1287/ijoo.2021.0069*

[31].   Filabadi, Milad Dehghani, and Sahar Pirooz Azad. "Robust optimisation framework for SCED problem in mixed AC-HVDC power systems with wind uncertainty." *IET Renewable Power Generation* vol. 14, no. 14, pp. 2563-2572, 2020, doi.org/10.1049/iet-rpg.2019.1127.

[32].   Dehghani Filabadi, Milad. "Robust optimization for SCED in AC-HVDC power systems". *University of Waterloo*, M.S. Thesis, 2019, doi: https://uwspace.uwaterloo.ca/handle/10012/14637

[33].   Filabadi, Milad Dehghani. "A New Paradigm in Addressing Data Uncertainty: Discussion and Future Research." *Academia Letters,* 2, 2022, doi.org/10.20935/AL4775

[34].   F. A. Simonen, "Pressure Vessels and Piping Systems: Reliability, Risk and Safety Assessment," p. 12.

[35].   T. Christopher, B. S. V. Rama Sarma, P. K. Govindan Potti, B. Nageswara Rao, and K. Sankarnarayanasamy, "A comparative study on failure pressure estimations of unflawed cylindrical vessels," *International Journal of Pressure Vessels and Piping*, vol. 79, no. 1, pp. 53–66, Jan. 2002, doi: 10.1016/S0308-0161(01)00126-0.

[36].   K. M. Rajan, P. U. Deshpande, and K. Narasimhan, "Experimental studies on bursting pressure of thin-walled flow formed pressure vessels," *Journal of Materials Processing Technology*, vol. 125–126, pp. 228–234, Sep. 2002, doi: 10.1016/S0924-0136(02)00298-4.

[37].   C. Zheng and S. Lei, "Research on bursting pressure formula of mild steel pressure vessel," *J. Zhejiang Univ. - Sci. A*, vol. 7, no. 2, pp. 277–281, Aug. 2006, doi: 10.1631/jzus.2006.AS0277.

[38].   M. Law and G. Bowie, "Prediction of failure strain and burst pressure in high yield-to-tensile strength ratio linepipe," *International Journal of Pressure Vessels and Piping*, vol. 84, no. 8, pp. 487–492, Aug. 2007, doi: 10.1016/j.ijpvp.2007.04.002.

[39].   A. Brabin, T. Christopher, and N. Rao, "Investigation on Failure Behavior of Unflawed Steel Cylindrical Pressure Vessels Using FEA," *Multidiscipline Modeling in Materials and Structures*, vol. 5, no. 1, pp. 29–42, Jan. 2009, doi: 10.1108/15736105200900002.

[40].   X.-K. Zhu and B. N. Leis, "Evaluation of burst pressure prediction models for line pipes," *International Journal of Pressure Vessels and Piping*, vol. 89, pp. 85–97, Jan. 2012, doi: 10.1016/j.ijpvp.2011.09.007.

[41].   N. Dwivedi, V. Kumar, A. Shrivastava, and R. Nareliya, "Burst Pressure Assessment of Pressure Vessel Using Finite Element Analysis: A Review," *J. Pressure Vessel Technol*, vol. 135, no. 4, Aug. 2013, doi: 10.1115/1.4023422.

[42].   J. Faupel, "Yield and bursting characteristics of heavy-wall cylinders," *Journal of Applied Mechanics*, vol. 23, pp. 1031–1064, 1956.

[43].   N. Svensson, "Bursting pressure of cylinderical and spherical vessels," *Journal of Applied Mechanics*, vol. 25, no. 80, pp. 89–96, 1958.

[44].   T. Christopher, B. S. V. Rama Sarma, P. K. Govindan Potti, B. Nageswara Rao, and K. Sankarnarayanasamy, "A comparative study on failure pressure estimations of unflawed cylindrical vessels," *International Journal of Pressure Vessels and Piping*, vol. 79, no. 1, pp. 53–66, Jan. 2002, doi: 10.1016/S0308-0161(01)00126-0.

[45].   C. Zheng and S. Lei, "Research on bursting pressure formula of mild steel pressure vessel," *J. Zhejiang Univ. - Sci. A*, vol. 7, no. 2, pp. 277–281, Aug. 2006, doi: 10.1631/jzus.2006.AS0277.

[46].   A. Brabin, N. Rao, and T. Christopher, "Investigation on Failure Behavior of Unflawed Steel Cylindrical Pressure Vessels Using FEA," *Multi Modelg in Mat & Struct*, vol. 5, no. 1, pp. 29–42, Jan. 2009, doi: 10.1108/15736105200900002.

[47].   Lee, Jeongwoo, Yanfeng Lu, Sumanth Kashyap, Aslan Alarmdari, Md Omar Faruk Emon, and Jae-Won Choi. "Liquid bridge microstereolithography." *Additive Manufacturing* 21 (2018): 76-83.

[48]. Alamdari, Aslan, Jeongwoo Lee, Myoeum Kim, Md Omar Faruk Emon, Ali Dhinojwala, and Jae-Won Choi. "Effects of surface energy reducing agents on adhesion force in liquid bridge microstereolithography." *Additive Manufacturing* 36 (2020): 101522.

[49]. Petersen, Shannon R., Jiayi Yu, Taylor R. Yeazel, Garrett Bass, Aslan Alamdari, and Matthew L. Becker. "Degradable, Photochemically Printable Poly (propylene fumarate)-Based ABA Triblock Elastomers." *Biomacromolecules* (2022).

[50]. Salama, Husien, Bandar Saman, Evan Heller, Raja Hari Gudlavalleti, Roman Mays, and Faquir Jain. "Twin Drain Quantum Well/Quantum Dot Channel Spatial Wave-Function Switched (SWS) FETs for Multi-Valued Logic and Compact DRAMs." *International Journal of High Speed Electronics and Systems* 27, no. 03n04 (2018): 1840024.

[51]. Gudlavalleti, R. H., B. Saman, R. Mays, H. Salama, Evan Heller, J. Chandy, and F. Jain. "A Novel Addressing Circuit for SWS-FET Based Multivalued Dynamic Random-Access Memory Array." In *Nanotechnology For Electronics, Biosensors, Additive Manufacturing And Emerging Systems Applications*, pp. 83-92. 2022.

[52]. Salama, H., B. Saman, R. H. Gudlavalleti, P. Y. Chan, R. Mays, B. Khan, E. Heller, J. Chandy, and F. Jain. "Simulation of Stacked Quantum Dot Channels SWS-FET Using Multi-FET ABM Modeling." In *NANOTECHNOLOGY FOR ELECTRONICS, PHOTONICS, BIOSENSORS, AND EMERGING TECHNOLOGIES*, pp. 129-133. 2021.

[53]. Salama, H., B. Saman, R. Gudlavalleti, R. Mays, E. Heller, J. Chandy, and F. Jain. "Compact 1-Bit Full Adder and 2-Bit SRAMs Using n-SWS-FETs." In *Nanotechnology For Electronics, Biosensors, Additive Manufacturing And Emerging Systems Applications*, pp. 129-138. 2022.

[54]. Salama, H., Bachr, C. "Implement of the information." In *International Journal of Innovation and Applied Studies,* 27 (2), pp. 456-462 2019.

[55]. Salama, H., Youssef, T. " Voltage Stability OF Transmission." In *International Journal of Innovation and Applied Studies,* 24 (2), pp. 439-445 2018.

[56]. Salama, H., B. Saman, R. Gudlavalleti, R. Mays, E. Heller, J. Chandy, and F. Jain. "Compact 1-Bit Full Adder and 2-Bit SRAMs Using n-SWS-FETs." In *Nanotechnology For Electronics, Biosensors, Additive Manufacturing And Emerging Systems Applications*, pp. 129-138. 2022.